\documentclass{osa-article}

\journal{osac}

\articletype{Research Article}
\usepackage{dsfont}
\usepackage{braket}

\begin{document}

\title{Exact calculation of stimulated emission driven by pulsed light}

\author{Kevin A. Fischer}

\address{E. L. Ginzton Laboratory, Stanford University, Stanford CA 94305, USA}
\email{kevinf@stanford.com}

\begin{abstract}
\normalsize Stimulated emission can be defined as the process when an incoming photon stimulates an additional quantum of energy from an atom into the same electromagnetic mode as the impinging photon. Hence, the two outgoing photons are identical. In a waveguide or free space, this intuition is typically found through Fermi's Golden rule, however, this does not properly account for the wave-like nature of the photons. Here, I present an exact solution to stimulated emission from a quantum two-level atom that properly accounts for the incoming and outgoing electromagnetic modes. This result is valid over a huge range of incident photon numbers. For a single incident photon, it shows how the photon must properly mode match the two-level system to cause stimulated emission. For a Fock state drive with large photon number, the exact solution shows how a two-level system Rabi oscillates with the traveling Fock mode as it passes by. I additionally use the same formalism to show that stimulated emission by a coherent pulse cannot be understood as an additional photon being emitted into the incident coherent state because the two-level system's state factorizes with the electromagnetic field's coherent state. Recent advances in superconducting circuits make them an ideal platform to test these predictions.
\end{abstract}

\section{Introduction}

Stimulated emission is one of the most fundamental effects in quantum optics, and is behind one of the most important inventions of the 20th century, the laser. In a laser, discrete electromagnetic modes representing a cavity interact with a gain medium consisting of few-level atoms. The atoms are incoherently pumped to their excited states and give their energy to the coherent field in the cavity modes via stimulated emission. This process has been rigorously studied and understood mathematically using non-perturbative density matrix theories \cite{carmichael2009statistical}.

In free space, structured materials, or waveguides, however, the dynamics of stimulated emission are an open topic of study. A common viewpoint, and the perspective that we will explore in this work, is that an incoming photon will stimulate emission from a two-level system prepared in its excited state, and cause the photon to be emitted into the same electromagnetic mode as the stimulating photon. To show this, most methods rely on perturbative approaches which involve considering the stimulated emission of a photon by $n$ photons, all with precise momentum $k$ and frequency $\lambda$ \cite{cohen1998atom,shankar2012principles}.

However, perturbative approaches to stimulated emission have limited explanatory power. For example, non-perturbative calculations have recently shown that a very specific incident photon shape is required to efficiently stimulate emission~\cite{valente2012optimal}, and stimulated effects considering pulse lengths can lead to phenomena such as optical diode rectification \cite{mascarenhas2016quantum}. Many approaches have now shown these types of results theoretically, however, often they are computationally intractable when more than a couple of photons stimulate the emission \cite{roy2017colloquium,pan2016exact,nysteen2015scattering,rephaeli2012stimulated}. Hence, they are incapable of recovering all experimental intuitions about stimulated emission. Further, as nanophotonics \cite{koenderink2015nanophotonics} brings realizations of these thought experiments closer to reality \cite{gu2017microwave,lodahl2017chiral,senellart2017high}, a better understanding of stimulated emission generally may be desired.

Here, I will develop a non-perturbative approach based on quantum stochastic calculus to build a fully accurate account of stimulated emission. Importantly, I show how to translate the central idea of the incoming and outgoing photons being identical into the context of photonic wavepackets. For this purpose, the most basic model is built around a two-level atom coupled to a bath of electromagnetic modes (schematic as Fig. 1), which could represent at its simplest a nanophotonic waveguide with single transverse spatial profile \cite{gardiner2004quantum}. I briefly note this causes no loss of generality in the model, which could easily be extended to more complicated electromagnetic baths so long as their couplings to the system are spectrally flat or in the white-noise limit \cite{gardiner1985cw,gardiner1992wave,li2017concepts}. 

\begin{figure}
\centering
\includegraphics[scale=1]{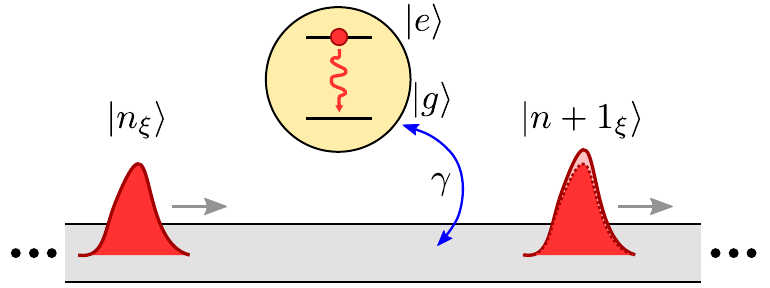}
  \caption{Illustration of stimulated emission from a two-level system, initially prepared in its excited state $\ket{e}$ and coupled to a unidirectional (chiral) waveguide at rate $\gamma$. A Fock state $\ket{n_\xi}$ has the potential to stimulate emission of a photon from a two-level system into the mode $\xi$. A coherent pulse driving the two-level atom cannot be thought of as stimulating emission into the same mode as the pulse.}
  \label{fig:1}
\end{figure}

The two-level atom has a ground $\ket{g}$ and excited state $\ket{e}$, and Hamiltonian
\begin{eqnarray}
H_\text{atom}=\omega_0 \sigma^\dag\sigma
\end{eqnarray}
with resonant frequency $\omega_0$ and dipole operator $\sigma=\ket{g}\bra{e}$. Meanwhile in the white-noise limit, the electromagnetic bath is described by 
\begin{eqnarray}
H_\text{EM}=\text{i}\int\mathop{\text{d}t} b^\dag(t)\frac{\partial}{\partial t} b(t)
\end{eqnarray}
where $b(t)$ is the continuous temporal-mode annihilation operator of the bath, with commutation $[b(t),b^\dag(s)]=\delta(t-s)$ and $b(t)\ket{\text{vac}}=0$. In the interaction picture, the total Hamiltonian is given by the dipolar energy exchange between the field and the atom at position $x=0$ where
\begin{eqnarray}
V(t)=\textrm{i}\sqrt{\gamma}\left(\sigma \tilde{b}^\dagger(t) -\sigma^\dagger \tilde{b}(t) \right);
\end{eqnarray}
$\gamma$ is the interaction rate and $\tilde{b}(t)=\text{e}^{\text{i}\omega_0 t}b(t)$.

With the white-noise limit, the Schr\"odinger equation for the unitary evolution operator $U(t)\ket{\psi(0)}\equiv U(t,0)\ket{\psi(0)}=\ket{\psi(t)}$ must be interpreted as a Stratonovich quantum stochastic differential equation (QSDE) \cite{gough2006quantum,gardiner2004quantum}
\begin{eqnarray}
\text{d}U(t)=-\text{i}V(t)\circ U(t).
\end{eqnarray}
(using $\hbar=1$). However, the Stratonovich QSDE is notoriously difficult to work with because it is defined as a midpoint integration. Conversion to It\=o form yields the more computationally helpful
\begin{equation}
\text{d}U(t)=\left(\sqrt{\gamma}\sigma\mathop{\text{d}B^\dag_t}-\sqrt{\gamma}\sigma^\dag\mathop{\text{d}B_t}-\frac{\gamma}{2}\sigma^\dag\sigma\mathop{\text{d}t}\right)U(t)
\end{equation}
where $\mathop{\text{d}B_t}=\tilde{b}(t)\mathop{\text{d}t}$ is the quantum noise increment. Importantly, the It\=o correction factor $-\frac{\gamma}{2}\sigma^\dag\sigma\mathop{\text{d}t}$ results in the noise increments commuting with past evolution operators $[\mathop{\text{d}B_t^\dag},U(s)]=[\mathop{\text{d}B_t},U(s)]=0$ for $s\leq t$. The It\=o QSDE can be formally integrated to find the evolution operator
\begin{equation}
U(t)=\mathcal{T}\exp\left(\int_0^t\big\{\sqrt{\gamma}\sigma\mathop{\text{d}B^\dag_t}-\sqrt{\gamma}\sigma^\dag\mathop{\text{d}B_t}-\frac{\gamma}{2}\sigma^\dag\sigma\mathop{\text{d}t}\big\}\right)\label{eq:Ut}
\end{equation}
and $\mathcal{T}$ indicates chronological ordering. Below, we use this QSDE to consider two scenarios: stimulated emission under Fock state drive and under coherent state drive.

\section{Fock state drive}

First, we consider the case where the two-level atom is initially prepared in the excited state $\ket{e}$, while the waveguide is prepared in a continuous-mode photon Fock state \cite{loudon2000quantum}
\begin{equation}
\ket{n_\xi}=\left(\int\mathop{\text{d}s}\xi(s)\tilde{b}^\dag(s)\right)^{n_\xi}\ket{\text{vac}}/\sqrt{n_\xi!}.
\end{equation}
Using the transformed mode operator $\tilde{b}(s)$ additionally re-centers the frequency of the photonic state around the natural resonance of the two-level system. Then, the temporal mode of the Fock state is $\xi$, which is normalized to $\int\mathop{\text{d}s}|\xi(s)|^2=1$. Notably, the continuous-mode Fock states are not unique, so for a different mode profile $\chi$ then $\braket{n_\chi|n_\xi}\neq 1$ necessarily. We denote the joint initial state of the system and waveguide $\ket{\psi_0}=\ket{n_\xi}\otimes\ket{e}\equiv\ket{n_\xi,e}$. Importantly, the It\=o table for Fock state drive remains unchanged from the vacuum one, i.e. $\mathop{\text{d}B_t}\mathop{\text{d}B_t^\dag}=\text{d}t$ and all other products involving two increments in the same time bin $[t,t+\text{d}t)$ are zero \cite{baragiola2012n}. Hence, the It\=o evolution Eq. \ref{eq:Ut} is still valid for the bath in a Fock state.

We are now prepared to mathematically translate the definition of stimulated emission: \emph{the probability the two-level system emits into the same mode as the incident Fock state}. This is equivalent to computing the overlap of the wavefunction with the final state $\ket{\psi_\text{stim}}=\ket{n+1_\xi,g}$ during and after emission. Specifically, we want to compute
\begin{eqnarray}
P_\text{stim}(t)=\left|\braket{n+1_\xi,g|U(t)|n_\xi,e}\right|^2.\label{eq:Pstim}
\end{eqnarray}
Because the initial and final states have definite photon number, the unitary propagator in Eq. \ref{eq:Pstim} can be expanded and grouped by number of system-waveguide scattering interactions involving a potential transfer of energy from $\sigma\mathop{\text{d}B^\dag_{t_{i}}}-\sigma^\dag{\text{d}B_{t_{i}}}$ to yield
\begin{subequations}
\begin{eqnarray}
\hspace{-15pt}P_\text{stim}(t)&=&|\bra{n+1_\xi,g}\Bigg[\sum_{p=1}^\infty\int_{0}^t\int_{t_1}^t\cdots\int_{t_{p-1}}^t\text{e}^{-\gamma\sigma^\dag\sigma t_p/2}\times\label{eq:expand}\\
&&\hspace{70pt}\overleftarrow{\prod\limits_{i=1}^{p}}\,\text{e}^{\gamma\sigma^\dag\sigma t_{i}/2}(\sqrt{\gamma}\sigma\mathop{\text{d}B^\dag_{t_{i}}}-\sqrt{\gamma}\sigma^\dag{\text{d}B_{t_{i}}})\text{e}^{-\gamma\sigma^\dag\sigma t_{i}/2}\Bigg]\ket{n_\xi,e}|^2\nonumber\\
&=&|\bra{n+1_\xi}\Bigg[\sum_{p=0}^{n_\xi}(-1)^{p}\int_{0}^t\int_{t_1}^t\cdots\int_{t_{2p}}^t\text{e}^{-\gamma t_{2p+1}/2}\sqrt{\gamma}\mathop{\text{d}B^\dag_{t_{2p+1}}}\times\nonumber\\
&&\hspace{70pt}\overleftarrow{\prod\limits_{i=1}^{p}}\text{e}^{\gamma(t_{2i}-t_{2i-1})/2}\gamma \mathop{\text{d}B_{t_{2i}}}\mathop{\text{d}B^\dag_{t_{2i-1}}}\Bigg]\ket{n_\xi}|^2\qquad\label{eq:intermediate}\\
&=&\left|\sum_{p=0}^{n_\xi}\left(-1\right)^p\sqrt{n_\xi+1}\frac{n_\xi!}{(n_\xi-p)!}F_p(t)\left(\sqrt{\gamma}\right)^{2p+1}\right|^2\label{eq:Pstim1}
\end{eqnarray}
\end{subequations}
where 
\begin{eqnarray}
&&\hspace{-20pt}F_p(t)=\int_{0}^t\int_{t_1}^t\cdots\int_{t_{2p}}^t\xi^*(t_{2p+1})\text{e}^{-\gamma t_{2p+1}/2}\mathop{\text{d}t_{2p+1}}\times\label{eq:Pstim2}\\
&&\hspace{45pt}\overleftarrow{\prod\limits_{i=1}^{p}}\xi(t_{2i})\xi^*(t_{2i-1})\text{e}^{\gamma(t_{2i}-t_{2i-1})/2}\mathop{\text{d}t_{2i}}\mathop{\text{d}t_{2i-1}},\nonumber
\end{eqnarray}
defining $\overleftarrow{\prod}_{i=k}^lg[i]\equiv\prod_{i=-l}^{-k}g[-i]$ and $t_0=0$. Moving from Eqs. \ref{eq:Pstim} to \ref{eq:expand} required only a re-expression of the unitary propagator and the orthonormality of the Fock states $\braket{n_\xi|m_\xi}=\delta_{nm}$, with the Kronecker-delta function. One of the key steps from Eqs. \ref{eq:expand} to \ref{eq:intermediate} is to understand how the nonlinearity of the two-level system requires an odd number of system-waveguide scattering interactions. Specifically, for the inner product between $\braket{g|A|e}\neq0$, the operator $A\propto\sigma$ or equivalently $A\propto\sigma(\sigma^\dag\sigma)^p$, which imposes the order of alternating emission $\text{d}B^\dag_{t_{2i}}$ and absorption $\text{d}B_{t_{2i-1}}$ scattering events. Here, also I mention that $\text{e}^{-\gamma\sigma^\dag\sigma t/2}\ket{g}=\ket{g}$ and $\text{e}^{-\gamma\sigma^\dag\sigma t/2}\ket{e}=\text{e}^{-\gamma t/2}\ket{e}$. Lastly, obtaining Eq. \ref{eq:Pstim1} requires the relation $\mathop{\text{d}B_t}\ket{n_\xi}=\mathop{\text{d}t}\sqrt{n_\xi}\xi(t)\ket{n-1_\xi}$, proved elsewhere \cite{baragiola2012n}. The expressions derived using these manipulations, Eqs. \ref{eq:Pstim1} and \ref{eq:Pstim2}, are a central result of this work and provide a complete solution to the stimulated emission problem.

\subsection{Exact evaluation of stimulated emission probability}

My next step is to show how these expressions for stimulated emission can be evaluated in practice. After the stimulating wavepacket has finished interacting with the system, the light stimulated into the mode $\xi$ keeps traveling along the waveguide uninterrupted. Hence, it will be convenient to work in the limit $S=\lim_{t\rightarrow\infty} U(t)$ where
\begin{equation}
P_\text{stim}=\left|\braket{n+1_\xi,g|S|n_\xi,e}\right|^2,
\end{equation}
since the final stimulated state is an eigenstate of $S$.

Now, I evaluate this expression for an explicit wavepacket. In particular, I use an exponentially decaying pulse shape with length $\tau$:
\begin{equation}
\xi(t)=\text{e}^{-t/2\tau}\Theta(t)/\sqrt{\tau},
\end{equation}
whose frequency is centered around the atomic resonance of the two-level system. ($\Theta$ is the Heaviside function.) This pulse was previously shown to be the optimal stimulating mode for a single photon when $\tau\approx0.35/\gamma$~\cite{valente2012optimal}. Then, there is a nice series expression for the coefficients
\begin{eqnarray}
\lim_{t\rightarrow\infty}F_p(t)=\frac{2^{p+1}}{p!}\frac{(\sqrt{\tau})^{2p+1}}{\prod_{k=0}^p (2k+1+\gamma\tau)}.
\end{eqnarray}
The resulting curves for $P_\text{stim}$ versus pulse length are plotted in Fig. 2 for several different photon numbers. Previously studies have looked at the lifetime of the two-level system to determine the optimal pulse length for maximum stimulation under single-photon drive \cite{valente2012optimal}, but here we can find this length directly from the probability to stimulate emission into the driving mode. I briefly note that the comparable definitions result in different conditions for optimal stimulation by a few percent, and the optimal projection can actually be onto a mildly different Fock mode than the drive (see Appendix A). As the number of photons increases the discrepancy decreases and, the probability to stimulate emission begins to Rabi oscillate due to a strong coupling between the driving field mode and the two-level system.

\begin{figure}
\centering
\includegraphics[scale=1]{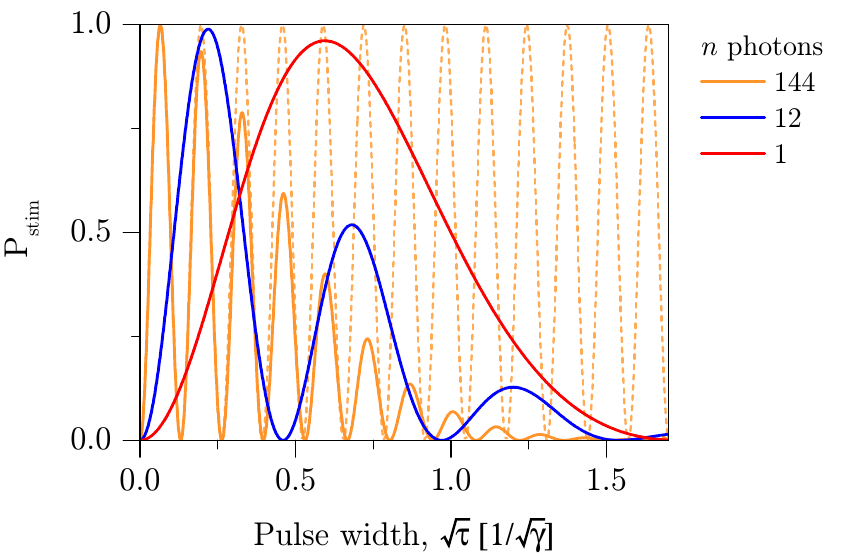}
  \caption{Probability for an $n$ photon Fock state in an exponentially decaying mode $\xi$ with length $\tau$ to stimulate emission from a quantum two-level system. Dashed line shows idealized model for large photon number and short pulses.}
  \label{fig:2}
\end{figure}

Further, for very large photon number
\begin{eqnarray}
\sqrt{n_\xi+1}\frac{n_\xi!}{(n_\xi-p)!}\approx \left(\sqrt{n_\xi}\right)^{2p+1}
\end{eqnarray}
and short pulses
\begin{eqnarray}
\lim_{t\rightarrow\infty}F_p(t)\approx\frac{(\sqrt{4\tau})^{2p+1}}{(2p+1)!}
\end{eqnarray}
for the dominant terms in the sequence, yielding an intuitive expression for the stimulated emission probability
\begin{eqnarray}
P_\text{stim}\approx\sin^2\left(\sqrt{4 \gamma\tau n_\xi}\right).\label{eq:PstimApprox}
\end{eqnarray}
This expression is plotted for 144 driving photons (orange dotted curve) in Fig. 2 and matches very well with the exact curve, though for longer pulses spontaneous emission spoils the ability to emit into the driving mode.

\subsection{Approximate scattered state}

Having evaluated the probability for stimulation, I now discuss the relation to the total final scattered state. In general, the final scattered state will have $n_\xi+1$ photons, but they may have a very complex entanglement. Previously, we calculated the projection onto all $n_\xi+1$ photons being in the mode $\xi$, which occurs with probability $P_\text{stim}$. Hence, the most general way to write the scattered state is given by
\begin{subequations}
\begin{eqnarray}
\ket{\Psi_\text{scatter}} &=&S\ket{n_\xi,e}\\
&=&\text{e}^{\text{i}\phi_\text{stim}}\sqrt{P_\text{stim}}\ket{\psi_\text{stim}} + \sqrt{P_0}\ket{\psi_\text{other}}
\end{eqnarray}
\end{subequations}
where $P_0$ is the probability no stimulation occurs and 
\begin{eqnarray}
&&\hspace{-20pt}\ket{\psi_\text{other}} =\\
&&\hspace{-20pt}\int\mathop{\text{d}s_1}\int\mathop{\text{d}s_2}\cdots\int\mathop{\text{d}s_{n_\xi+1}}\zeta(s_1,s_2,\dots,s_{n_\xi+1})\tilde{b}^\dag(s_1)\tilde{b}^\dag(s_2)\cdots\tilde{b}^\dag(s_{n_\xi+1})\ket{\text{vac},g}/\sqrt{(n_\xi+1)!}\nonumber
\end{eqnarray}
is the resulting (and normalized) wavefunction. In principle, it is possible to solve for the total wavefunction $\ket{\Psi_\text{scatter}}$ \cite{xu2015input}, but it is intractable when many photons interact with the two-level system.

We can, however, consider the case of very short pulses, which yields the simple expression
\begin{eqnarray}
\ket{\psi_\text{other}}\approx\sqrt{\gamma}\int\mathop{\text{d}s}\text{e}^{-\gamma s/2}b^\dag(s)\ket{n_\xi,g}/\sqrt{n_\xi+1}\label{eq:decay}
\end{eqnarray}
(or equivalently $\zeta(s_1,s_2,\dots,s_{n_\xi+1})=\sqrt{\gamma}\text{e}^{-\gamma s_1/2}\xi(s_2)\cdots\xi(s_{n_\xi+1})$).
This physically can be understood as the incident pulse briefly Rabi flopping with the two-level system and then the remaining excitation of the system decaying spontaneously. For very short pulses with a large number of photons, it is also possible to show that
\begin{subequations}
\begin{eqnarray}
P_\text{0}&=&\left|\braket{n_\xi,e|U(T)|n_\xi,e}\right|^2\\
&\approx&\cos^2\left(\sqrt{4 \gamma\tau n_\xi}\right)\label{eq:P0Approx}
\end{eqnarray}
\end{subequations}
of the population is not stimulated and the field is left in its initial state immediately after the system-pulse interaction (but before spontaneous emission has time to occur, i.e. $\tau\ll T\ll 1/\gamma$). This population subsequently decays spontaneously with the standard exponential, yielding the state in Eq. \ref{eq:decay}.

\subsection{Rabi oscillations between the field and system}

Previously, we have described the stimulated emission interaction in terms of Rabi oscillations between the traveling Fock mode and the two-level system, which we inferred based on the post interaction oscillations. In this section, I explicitly show the oscillations that occur during the pulse-system interaction.

To obtain a tractable solution for the exponential pulse shape, I assume the short pulse limit, where the exponential decay from the It\=o correction factor can be ignored. This yields the relatively simple closed solution with
\begin{eqnarray}
F_p(t)\approx\frac{\sqrt{4\tau}^{2p+1}}{(2p+1)!}\text{e}^{-\frac{2p+1}{2\tau}t}\left(-1+\text{e}^{t/{2\tau}}\right)^{2p+1}
\end{eqnarray}
for the projection onto $\ket{n+1_\xi,g}$. Because we want to see Rabi oscillations between $\ket{n_\xi,e}$ and $\ket{n+1_\xi,g}$ while the pulse interacts with the system, I additionally provide the solution for the projection of the wavefunction onto $\ket{n_\xi,e}$ :
\begin{subequations}
\begin{eqnarray}
P_0(t)&=&\left|\braket{n_\xi,e|U(t)|n_\xi,e}\right|^2\\
&=&\left|\sum_{p=0}^{n_\xi}\left(-1\right)^p\frac{n_\xi!}{(n_\xi-p)!}G_p(t)\left(\sqrt{\gamma}\right)^{2p}\right|^2
\end{eqnarray}
\end{subequations}
where
\begin{eqnarray}
G_p(t)\approx\frac{\sqrt{4\tau}^{2p}}{(2p)!}\text{e}^{-\frac{2p}{2\tau}t}\left(-1+\text{e}^{t/{2\tau}}\right)^{2p}.
\end{eqnarray}
Notably, when $t\gg\tau$ such that the entire pulse has interacted with the system, these solutions reduce to Eqs. \ref{eq:PstimApprox} and \ref{eq:P0Approx}.

\begin{figure}
\centering
\includegraphics[scale=1]{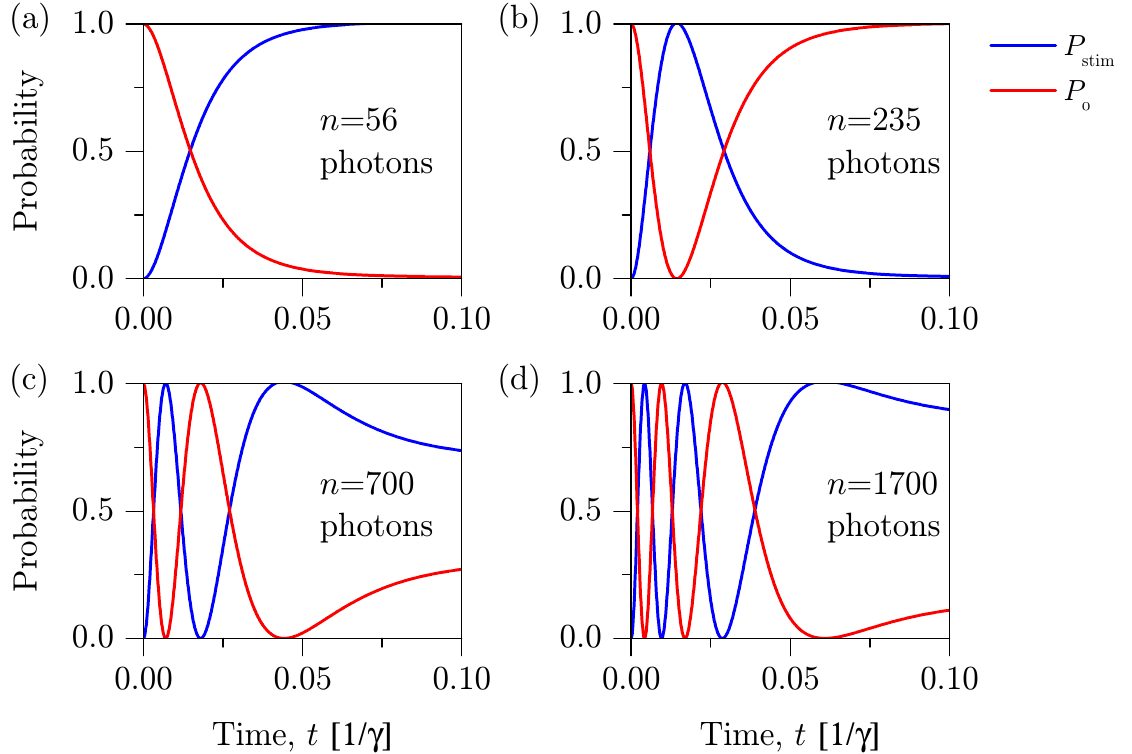}
  \caption{Rabi oscillations of a quantum between the two-level system and a stimulated Fock mode, as seen in the probabilities $P_\text{stim}(t)=\left|\braket{n+1_\xi,g|U(t)|n_\xi,e}\right|^2$ and $P_0(t)=\left|\braket{n_\xi,e|U(t)|n_\xi,e}\right|^2$. The driving Fock mode has a pulse width $\tau=0.01/\gamma$. (a)-(d) show drive by Fock states of different photon number.}
  \label{fig:3}
\end{figure}

The projections for $\ket{n_\xi,e}$ and $\ket{n+1_\xi,g}$ clearly show coherent oscillation between the quantum initially in the two-level system and an addition photon in the incident Fock mode $\xi$ (see Fig. \ref{fig:3}). Because these projections form a complete basis for the pure state in the short pulse limit, during the energy exchange the system and field become maximally entangled. Lastly, I note that the period of oscillation decreases nonlinearly with time because the energy density of the incident Fock state is also decreasing due to the exponential envelope (a square pulse provides constant frequency oscillations, see Appendix B).

This is the first definitive proof to my understanding, that when a two-level system is driven by a short pulse it is possible to stimulate oscillations between the bath mode $\xi$ \textit{identically} and the two-level system. I find the result quite remarkable, because the two-level system only ever exchanges one quantum of energy at a time with the waveguide during each bin $[t,t+\text{d}t)$, but it does so in a completely coherent manner through different interference pathways that leads to emission of an additional photon into the mode $\xi$. These results have been recently suggested by an interesting technique of Fock-state master equations \cite{baragiola2012n}, however, the calculations trace over the past field states and hence could not definitely conclude stimulation into the exact same field mode $\xi$.

\section{Coherent state drive}

A coherent state has long been known to stimulate emission from a two-level atom, using definitions such as radiation-induced emission (introduced by Einstein in 1917), as a negative absorption \cite{mollow1972stimulated}, and applied for a single atom in a one-dimensional environment \cite{valente2012monitoring}. In this section, I briefly summarize how the definition proposed here applies to a coherent state, i.e. the concept of stimulated emission into the mode of the incident field. The two-level atom is initially prepared in its the excited state $\ket{e}$, while the waveguide is prepared in a continuous-mode coherent state
\begin{equation}
\ket{\alpha}=\exp\left(\int \{\alpha(s)\tilde{b}^\dag(s) - \alpha^*(s)\tilde{b}(s)\}\mathop{\text{d} s} \right)\ket{\text{vac}},
\end{equation}
which we write as $\ket{\psi_0}=\ket{\alpha}\otimes\ket{e}$. As with Fock state drive, the It\=o algebra for coherent state drive remains unchanged from the vacuum one, so the It\=o evolution equation can again be used. This point can be proved similarly as the Fock state It\=o table using the relation 
\begin{equation}
\mathop{\text{d}B_t}\ket{\alpha}=\mathop{\text{d}t}\alpha(t)\ket{\alpha}.\label{eq:coh}
\end{equation}

For coherent drive, the authors in Refs. \cite{fischer2017signatures} have shown that stimulated emission takes a very different form for coherent drive. In a similar manner as our Fock state example, we will compute the overlap with a final coherent state
$\ket{\beta}$ during and after emission. Specifically, we want to compute
\begin{equation}
\braket{\beta|\psi(t)}=\braket{\beta|U(t)|\alpha}\otimes\ket{e}
\end{equation}
Using Eq. \ref{eq:coh} we can simplify this expression
\begin{equation}\textstyle
\braket{\beta|\psi(t)}=\mathcal{T}\text{e}^{\int_0^t\{\sqrt{\gamma}\beta^*(t)\sigma-\sqrt{\gamma}\alpha(t)\sigma^\dag-\frac{\gamma}{2}\sigma^\dag\sigma\}\mathop{\text{d}t}}\ket{e}\braket{\beta|\alpha}.
\end{equation}
There are three immediately obvious features from this expression:
\begin{enumerate}
\item As the pulse becomes longer, the It\=o correction factor again causes the photons to be emitted into a bunch of modes other than in the initial state.
\item The probability of oscillation between ground and excited states is maximized by the choice of $\beta=\alpha$. 
\item The projection completely factorizes into an atom part and a coherent field part.
\end{enumerate}
Therefore, the coherent drive cannot be considered to stimulate a quantum of energy from the two-level system into any single outgoing mode of a coherent state. Instead, the field remains mostly un-entangled with the system while the system undergoes Rabi oscillations. The excited two-level system of course must emit its energy, but it does so in a somewhat unusual fashion involving multiple photon emissions \cite{fischer2017signatures}. In the short pulse limit for a square mode profile with frequency centered around the natural resonance of the two-level system $P_e(t)=\left|\braket{\alpha,e|\psi(t)}\right|^2=\cos^2(\sqrt{\gamma}|\alpha|t)$ and $P_g(t)=\left|\braket{\alpha,g|\psi(t)}\right|^2=\sin^2(\sqrt{\gamma}|\alpha|t)$ during $0\leq t\leq T$, which are the standard Rabi oscillations of a two-level system under coherent state drive.

\section{Conclusions}

In summary, I have shown how to use quantum stochastic calculus to solve the stimulated emission problem exactly. In particular, under drive by a continuous-mode Fock state the interaction with a two-level system has a particularly remarkable behavior in the short pulse and large photon number regime. Despite only exchanging one quantum of energy with the bath sequentially, the collective behavior could be modeled as an oscillating Jaynes--Cummings system given a cavity mode prepared with $n$ photons. On the other hand, stimulated emission by a coherent pulse in free-space or a waveguide channel cannot be understood as an additional photon being emitted into the incident coherent state. This results from the factorizability of the waveguide coherent states with the system state at all times. In contrast, a two-level system coupled to a single cavity mode prepared in a coherent state quickly produces maximal atom-cavity entanglement \cite{agarwal1991nonclassical}.

Devices based on superconducting transmons have already shown the ability to generate tunable traveling Fock or coherent states \cite{heeres2017implementing,axline2018demand} and to act as high-quality two-level systems for nonclassical light generation \cite{lang2011observation,lang2013correlations}. Therefore, they are an ideal platform for testing the theoretical predictions put forth in this letter. Physically, the probability of stimulation could be measured using the dynamic state transfer method, whereby the photons in a traveling Fock mode of interest are transferred to a cavity---state tomography is then performed on the cavity \cite{axline2018demand}.

In a quantum circuit with ideal behavior, stimulated emission could be used in a chain of qubits to sequentially grow Fock states. As the Fock state grew, each qubit would need to have its coupling tuned slower to keep the rotation angle fixed so that only one Rabi flop occurs per stimulation, where $P_\text{stim}$ is always near unity. This chain of qubits with decreasing coupling could be implemented with the superconducting tunable coupling qubit (TCQ) \cite{gambetta2011superconducting}. Starting from just a few photons, a highly-pure Fock state of large photon number could be generated. This process could even potentially be used to create NOON states with huge photon number for quantum-limited metrology~\cite{giovannetti2011advances}.

\section{Acknowledgements}

I thank Jelena Vu\v{c}kovi\'c for helpful feedback and discussions, and I also acknowledge helpful discussions with David AB Miller, Vinay Ramasesh, Joshua Combes, Daniil Lukin, and Rahul Trivedi, as well financial support from the Air Force Office of Scientific Research (AFOSR) MURI Center for Quantum Metaphotonics and Metamaterials.

\appendix

\section{Optimal projection}

\begin{figure}[b]
\centering
\includegraphics[scale=1]{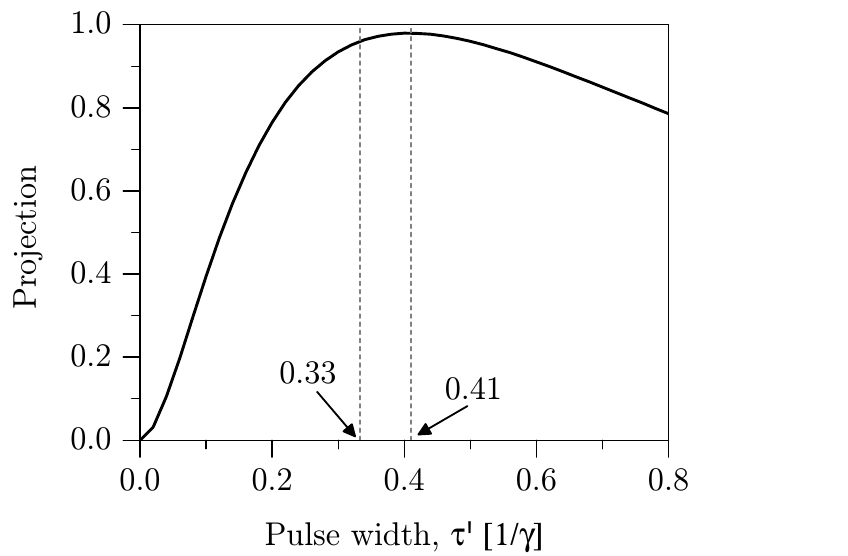}
  \caption{We defined the probability of stimulated emission as $P_\text{stim}=\left|\braket{n+1_\xi,g|S|n_\xi,e}\right|^2$, however, the projection onto a different Fock mode than the incident one could potentially be larger. Here, we consider $\text{Projection}=\left|\braket{n+1_{\xi,\tau'},g|S|n_{\xi,\tau},e}\right|^2$ where the state is driven by a pulse of width $\tau=1/3\gamma$, but the pulse width $\tau'$ of the state projected onto is variable.}
  \label{fig:4}
\end{figure}

As mentioned in the main text, optimal stimulated emission need not strictly occur into the incident Fock mode. Here, I explore this potential discrepancy by considering the exact scattered solution for stimulated emission by a single photon. When $\tau=1/3\gamma$ \cite{pan2016exact}, the projection of the exact scattered photonic state onto the temporal modes is given by
\begin{eqnarray}
\Psi_\text{scatter}(\tau_1,\tau_2)=\gamma\sqrt{3}\left(\text{e}^{-\frac{3\gamma \tau_1+\gamma \tau_2}{2}}\Theta(0\leq \tau_2\leq \tau_1)+\text{e}^{-\frac{3\gamma \tau_2+\gamma \tau_1}{2}}\Theta(0\leq \tau_1< \tau_2)\right).
\end{eqnarray}
The scattered state can be projected onto the two-photon Fock state of width $\tau'$ using
\begin{eqnarray}
\psi_\text{Fock}(s_1,s_2)=\frac{1}{\tau'\sqrt{2}}\text{e}^{-\frac{s_1+s_2}{2\tau'}}\Theta(s_1)\Theta(s_2)
\end{eqnarray}
and the overlap
\begin{subequations}
\begin{eqnarray}
\text{Projection}&=& \left|\braket{\psi_\text{Fock}|\Psi_\text{scatter}}\right|^2\\
&=&\int\mathop{\text{d}s_1}\int\mathop{\text{d}s_2}\left(\psi_\text{Fock}^*(s_1,s_2)\Psi_\text{scatter}(s_1,s_2) + \psi_\text{Fock}^*(s_1,s_2)\Psi_\text{scatter}(s_2,s_1) \right).\qquad
\end{eqnarray}
\end{subequations}
I numerically evaluated this integral and plotted the results in Fig. \ref{fig:4}. As expected, the projection is small when there is a large mismatch between the stimulating pulse width and the projected pulse width. However, the projection is not optimal when the pulse widths are exactly equal---it is maximized when $\tau'$ is slightly larger than $\tau$. Further investigation of this deviation might be interesting, exploring the optimal Fock mode to project onto for single-photon stimulation.

Nevertheless, this small discrepancy decreases further for short pulses and large photon number, as evidenced by the stimulation probabilities trending towards unity in Fig. \ref{fig:2}. Hence, we largely ignore its effects in this work.

\section{Drive by a square Fock mode}

In the main text, I used the exponential pulse because it yields an exact yet simple expression for stimulated emission. Drive by a square Fock mode \begin{eqnarray}
\xi(t)=1/\sqrt{T}\Theta(0\leq t<T),
\end{eqnarray}
where $T$ is the width of the square pulse, does not yield such a nice general expression. It does, however, give a convenient solution in the short pulse/high photon number limit. This result helps us to understand the Rabi oscillations that occur between the Fock mode and the two-level system. Specifically, applying the methods discussed for this new mode profile in the short pulse/high photon number limit gives
\begin{eqnarray}
P_\text{stim}(t)\approx\sin^2\left(t\sqrt{\tfrac{\gamma }{T}n_\xi}\right)\Theta(0\leq t<T)+\sin^2\left(\sqrt{\gamma T n_\xi}\right)\Theta(T\leq t)
\end{eqnarray}
and
\begin{eqnarray}
P_0(t)\approx\cos^2\left(t\sqrt{\tfrac{\gamma}{T}n_\xi}\right)\Theta(0\leq t<T)+\cos^2\left(\sqrt{\gamma T n_\xi}\right)\Theta(T\leq t).
\end{eqnarray}
Hence, stimulated emission can be viewed as a coherent interaction at rate $g_\text{eff}=\sqrt{\gamma/T}$ between the two-level system and a single bosonic mode at Rabi frequency $\omega_\text{R}=g_\text{eff}\sqrt{n_\xi}$ for a brief period of time, as was suggested in Ref. \cite{silberfarb2003continuous}. As a brief aside, changing the initial condition to, e.g. absorption of the Fock state rather than stimulated emission, results in similar solutions for the probability of pulse transparency versus absorption.

\bibliography{bibliography}



\end{document}